\documentclass[journal]{IEEEtran}
 \usepackage{graphicx}

\ifCLASSINFOpdf
\else
\fi
\begin{document}
%
\title{Distributed Demand Response and User Adaptation in Smart Grids}
%
%
%

\author{Zhong Fan,~\IEEEmembership{Senior Member,~IEEE}
\thanks{Z. Fan is with Toshiba Research Europe Limited,
Telecommunications Research Laboratory, 32 Queen Square, Bristol,
BS1 4ND, UK. e-mail: zhong.fan@toshiba-trel.com}}

\maketitle

\begin{abstract}
This paper proposes a distributed framework for demand response and
user adaptation in smart grid networks. In particular, we borrow the
concept of congestion pricing in Internet traffic control and show
that pricing information is very useful to regulate user demand and
hence balance network load. User preference is modeled as a
willingness to pay parameter which can be seen as an indicator of
differential quality of service. Both analysis and simulation
results are presented to demonstrate the dynamics and convergence
behavior of the algorithm.
\end{abstract}


%
\IEEEpeerreviewmaketitle

\section{Introduction}

A smart grid is an {\it intelligent} electricity network that
integrates the actions of all users connected to it and makes use of
advanced information, control, and communication technologies to
save energy, reduce cost and increase reliability and transparency.
Recently, many countries have started massive efforts on research
and developing smart grids. For example, the smart grid is a vital
component of President Obama's comprehensive energy plan: the
American Recovery and Investment Act includes 11 billion USD in
investments to ``jump start the transformation to a bigger, better,
smarter grid''.

In electricity grids, demand response (DR) is a mechanism for
achieving energy efficiency through managing customer consumption of
electricity in response to supply conditions, e.g., having end users
reduce their demand at critical times or in response to market
prices. In the future smart grid, the two way communications between
energy provider and end users enabled by advanced communication
infrastructure (e.g., wireless sensor networks and power line
communications) and protocols will greatly enhance demand response
capabilities of the whole system. In contrast to the current simple
time-of-use (TOU) pricing (e.g. peak time vs. off-peak time), it can
be envisaged that a more dynamic, real-time adaptation to market
prices would not only enable consumers to save more energy and
money, as well as manage their usage preferences more flexibly, but
also facilitate the grid move closer towards its optimal operating
point. For a recent overview of challenges and issues of enabling
communication technologies in this area, please refer to
\cite{fan10}. The authors of \cite{rahimi} argue that demand
response and distributed energy storage can be seen as distributed
energy resources and are main drivers of smart grid. While DR can
help the industry to achieve market efficiency and operational
reliability, there are also challenges ahead in implementing DR
under smart grid and market paradigms.

There are a few papers recently on smart grid DR using load
scheduling. In \cite{wangchen}, user preferences are taken into
account with the concept of discomfort level and an optimization
problem is formulated to balance the load and minimize the user
inconvenience caused by demand scheduling. Several ideas from the
distributed computing area such as makespan have been introduced to
energy consumption optimization. Similarly, in \cite{wong}, an
energy consumption scheduling problem is established to minimize the
overall energy cost. Techniques similar to those used in wireless
network resource allocation have been applied here to solve the
underlying optimization problem. In both works, the user demands are
known beforehand and the optimization problem is solved in numerical
iterations.

In this paper, we consider a fully distributed system where the only
information available to the end users is the current price which is
dependent on the overall system load. Based on this information, the
users try to adapt their demands so as to maximize their own
utility. There is no central control entity. Inspired by the
well-established work on congestion pricing in IP networks, we
propose a simple adaptation strategy based on price feedbacks and
show that it is very effective in achieving demand response.

The rest of the paper is organized as follows. Section 2 introduces
our DR model and the adaptation algorithm. We present some
simulation results in Section 3. Conclusions and future work are
presented in Section 4.

\section{Demand response model}

\subsection{Congestion pricing background}
In this paper we propose to apply the principle of congestion
pricing in IP networks to demand response in the electricity grid.
In their seminal paper \cite{kelly}, Kelly {\it et al.} have
proposed the proportionally fair pricing (PFP) scheme in which each
user declares a price per unit time that he is willing to pay for
his flow. In that sense the network capacity is shared among the
flows of all users in proportion to the prices paid by the users. It
has been shown in \cite{kelly} that in a weighted proportionally
fair system where the weights are the prices the users pay per unit
time, when each user chooses the price that maximizes the utility
she gets from the network, the system converges to a state where the
total utility of the network is maximized. In other words, in an
ideal environment, the PFP proposal is able to decentralize the
global optimal allocation of congestible resources. Another
important result of \cite{kelly} is that rate control (such as TCP)
based on additive increase and multiplicative decrease achieves
proportional fairness. It has been proved in \cite{glen} that the
decentralized congestion control mechanism is stable even under
arbitrary network topologies and heterogeneous round trip times
(feedback delays).

In Kelly's approach, the philosophy is that users who are willing to
pay more should get more. As the network makes no explicit promises
to the user, there is no need for over provisioning in the core of
the network. One implementation of PFP is to give control to end
systems (users). In this scheme, the TCP algorithm is modified to
incorporate congestion prices by means of protocols like explicit
congestion notification (ECN) \cite{ecn}. Upon receiving feedback
signals, $f(t)$, which are related to shadow prices (in terms of
packet marks), the users are free to react as they choose, but will
incur charges when resources are congested. An end system can adjust
its rate $x(t)$ using a willingness to pay (WTP) parameter $w$:
\begin{equation}
x(t + 1) = x(t) + \alpha(w - f(t)),
\end{equation}
where $\alpha$ affects the rate of convergence of the algorithm.

In \cite{ganesh}, explicit prices instead of marks are fed back to
the end users as incentives and users adapt their rates accordingly.
It has been shown that the system converges to an optimal allocation
of bandwidth: the users' price predictions converge to the actual
price and their bandwidth allocations converge to levels which
equalize their marginal utility of bandwidth to the price of
bandwidth.

\subsection{The DR model and user adaptation}

We consider a discrete time slot system where $N$ users share some
energy resources. In each time slot $n$, user $i$ has a demand of
$x_i(n)$ (e.g. hourly energy consumption if the time granularity is
one hour). The unit price of energy in a time slot is a function of
the aggregate demand:
\begin{equation}
p(n) = f(\sum_{i = 1}^N x_i(n)).
\end{equation}
The price function (spot market price) can be of the following form
\cite{wangchen}\cite{lemmon}:
\begin{equation}
f(x) = a({x \over C})^k, \label{a1}
\end{equation}
where $a$ and $k$ are constants, and $C$ is the capacity of the
market.

Each user $i$ is associated with a utility function $u_i(x_i(n))$ in
time slot $n$, which is a concave, non-decreasing function of its
demand. A typical logarithmic utility function is given by
\cite{kelly}:
\begin{equation}
u_i(x) = w_i \log x,
\label{a2}
\end{equation}
where $w_i$ is the willingness to pay parameter. Hence user $i$
chooses its demand $x_i(n)$ to seek to maximize
\begin{equation}
u_i(x_i(n)) - x_i(n)p(n).
\label{a3}
\end{equation}

We would like to elaborate on a few assumptions made in the above
model. Firstly, in our model the demand $x$ is a continuous
variable, which may not be realistic in practice. For example, in
real life, the daily usages of a washing machine and a dryer are
(fixed as) 1.49 kWh and 2.50 kWh respectively. Here the adaptation
of $x$ can be seen as an action of load scheduling: for example,
re-scheduling a dryer operation from time slot $n = 1$ to slot $n =
3$ leads to $x(1)$ reduced by 2.50 kW per hour and $x(3)$ increased
by 2.50 kW per hour. Secondly, how to characterize user preference
is an open issue. For instance, a user may prefer his washing done
at 6pm which is a typical peak time. To some extent, this preference
can be reflected in the WTP parameter $w$ in (\ref{a2}): when a user
is willing to pay more, he/she can have a higher demand. However,
the delay (or waiting time) incurred due to rescheduling is not
considered in this model. Thirdly, as pointed out in \cite{kelly},
logarithmic utility functions lead to proportional fairness. There
are other types of utility functions available corresponding to
different fairness criteria, e.g. $u_i(x) = w_i{{x^{\beta} - 1}
\over \beta}, \beta < 1$ as proposed in \cite{ganesh}. How to choose
a most suitable utility function in DR applications is an open
issue, e.g. how to factor in the waiting time and user discomfort
level \cite{wangchen}.

User $i$ adapts its demand according to the following equation:
\begin{equation}
x_i(n + 1) = x_i(n) + \alpha_i(w_i - x_i(n)p(n)), \label{a4}
\end{equation}
where $\alpha_i$ is a parameter that controls the rate of
convergence of the algorithm. It is clear that the user adjusts her
demand according to the price information ($p(n)$) and her own
willingness to pay preference ($w$).


To show that the above adaptation converges to the user optimum, let
us assume that the equilibrium price is $q$. Then by solving
$u'_i(x_i(n)) = q$, we have the optimal demand $x_i^*$ as
\begin{equation}
x_i^* = {w_i \over q}. \label{a6}
\end{equation}
Given (\ref{a4}), the error of demand estimate, $e_i(n + 1)$, is
given by
\begin{equation}
e_i(n + 1) = x_i(n + 1) - x_i^* = x_i(n) + \alpha_i(w_i - x_i(n)q)-
{w_i \over q}. \label{a7}
\end{equation}
Then it follows that
\begin{equation} e_i(n + 1) = (1 - \alpha_i
q)(x_i(n) - {w_i \over q}) = (1 - \alpha_i q)e_i(n).
\end{equation}
Therefore $e_i(n)$ is a geometric series, and when $|1 - \alpha_i q|
<1, \lim_{n \rightarrow \infty} e_i(n) = 0$. This has established
that with properly chosen $\alpha_i$, the adaptation will converge
to the optimum. Following \cite{kelly}, it is also straightforward
to establish the global stability of the algorithm in a differential
equation form (\ref{ee}) using an appropriate Lyapunov function.
\begin{equation}
{d \over {dt}}x_i(t) = \alpha_i(w_i - x_i(t)p(t)). \label{ee}
\end{equation}

\subsection{Implementation considerations}

\begin{figure}
\vspace{-30mm}
 \centering
   \includegraphics[height=6.0in]{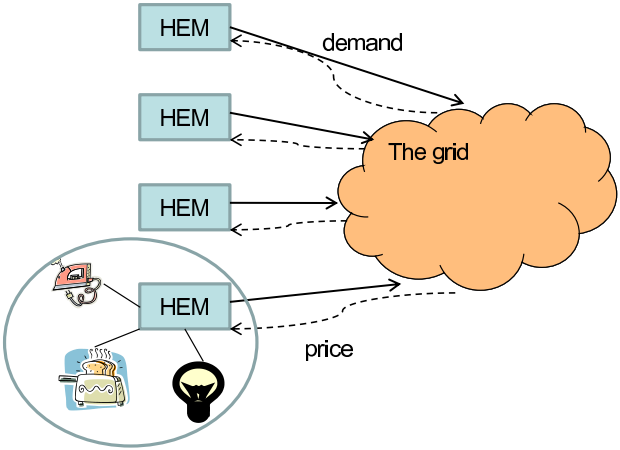}
   \vspace{-55mm}
\caption{Distributed demand response}

   \label{hem}
\end{figure}

In a residential energy management scenario, we envisage that each
user in our model is represented by an entity or software agent
called home energy manager (HEM) at a consumer's home. Appliances in
the home are equipped with smart meters, and they communicate with
HEM via low power wireless such as ZigBee. HEM is further connected
to the grid (supplier) via either wired or wireless links. Based on
the price information it receives, HEM calculates demand in the next
time slot and distributes it to different appliances. The overall
architecture is shown in Figure \ref{hem}.

We note that some appliances like refrigerator and heating have {\it
hard} consumption scheduling requirements, while others such as
washing machine have {\it soft} requirements \cite{wong}. When HEM
has to shift the demand to another time slot, it may apply only to
soft appliances. For example, HEM obtains $x(n + 1)$ based on
(\ref{a4}) with WTP parameter $w$. If it can satisfy the demand from
the hard appliances (denoted by $h$), it can re-schedule the demand
from some of the soft appliances (denoted by $s$) so that $x(n + 1)
= h(n + 1) + s(n + 1)$. On the other hand, if it cannot meet the
demand from the hard appliances, HEM may have to increase $w$ and
recalculate $x(n + 1)$.

\section{Simulation results}

In this section, we use simulations to study the behavior and
dynamics of the proposed algorithm. There are $N = 10$ users and
without loss of generality we assume that the capacity $C$ is 1. For
the price function (\ref{a1}), $a = 1, k = 4$.

\subsection{Basic simulation}
\label{s1}

\begin{figure}
\centering
\includegraphics[width=3.0in]{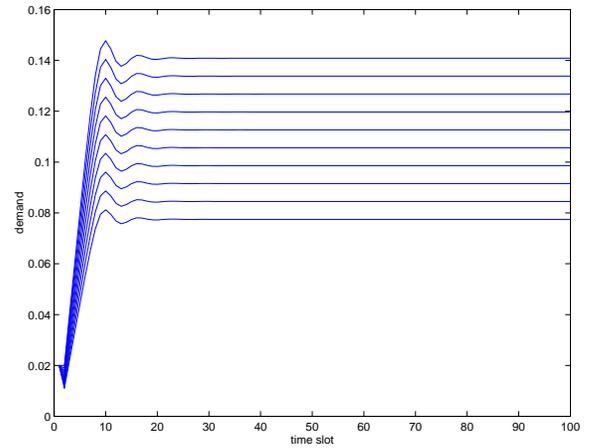}
 \caption{Simulation \ref{s1}: demand adaptation of 10 users}

   \label{demand1}
\end{figure}

\begin{figure}
\centering
\includegraphics[width=3.0in]{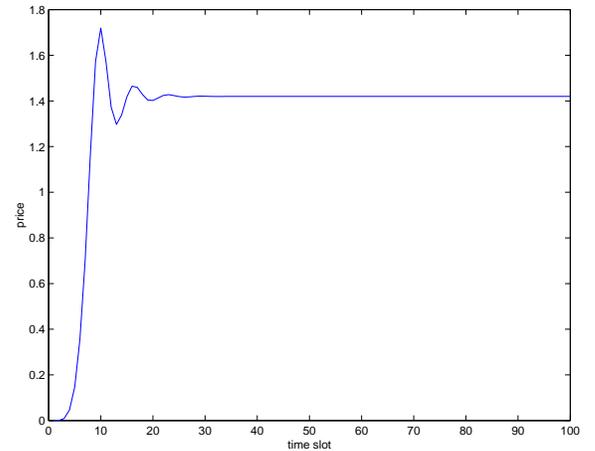}
 \caption{Simulation \ref{s1}: price evolution}
   \label{price1}
\end{figure}

Here all the users initiate their demands at 0.02, and their
willingness to pay parameters range from 0.11 (user 1) to 0.20 (user
10). All the users have the same adaptation parameter $\alpha$ of
0.1. Figure \ref{demand1} shows the demand changes with time for 10
users. After a short transient period, each user demand converges to
a stable value (determined by different $w$ values). It is also
evident that $w$ is a crucial factor in determining how aggressive a
user should be responding to the price signals. Figure \ref{price1}
clearly shows that the price converges to the optimal value. When
the system reaches its equilibrium (assuming $a = 1, C = 1$), we
have
\begin{equation}
x_i =  {w_i \over {({\sum_{i = 1}^N x_i(n)})^k}}. \label{b1}
\end{equation}
Summing over $i$ on both sides of (\ref{b1}), it is easy to verify
that the price at equilibrium is
\begin{equation}
p =  ({\sum_{i = 1}^N w_i(n)})^{k \over {k + 1}}.
\end{equation}
In this case, $p = 1.42$ as shown in Figure \ref{price1}.

\subsection{The effect of $\alpha$}
\label{s2}

\begin{figure}
\centering
\includegraphics[width=3.0in]{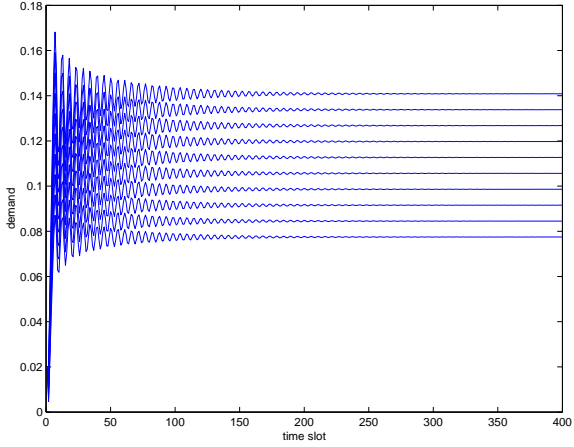}
 \caption{Simulation \ref{s2}: demand adaptation of 10 users}

   \label{demand2}
\end{figure}

\begin{figure}
\centering
\includegraphics[width=3.0in]{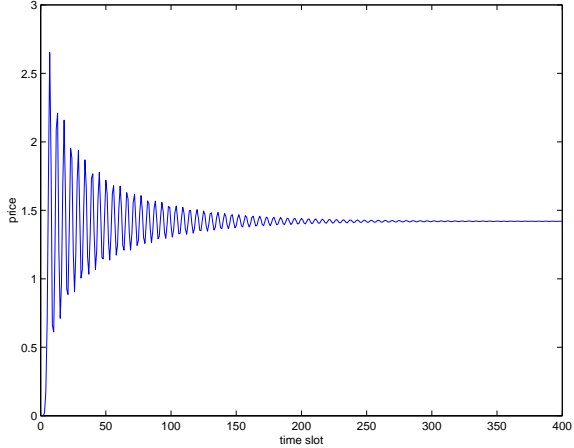}
 \caption{Simulation \ref{s2}: price evolution}
   \label{price2}
\end{figure}

In this simulation experiment we study the effect of $\alpha$ on
system performance. Figure \ref{demand2} and Figure \ref{price2}
depict the demand and price evolution versus time respectively for
$\alpha  = 0.17$. Compared with Figure \ref{demand1} and Figure
\ref{price1}, it can be seen that with a larger $\alpha$, it takes
much longer to converge. Therefore $\alpha$ is an important system
parameter that controls the convergence speed of the process.

\subsection{Heterogeneous initial demands}
\label{s3}

\begin{figure}
\centering
\includegraphics[width=3.0in]{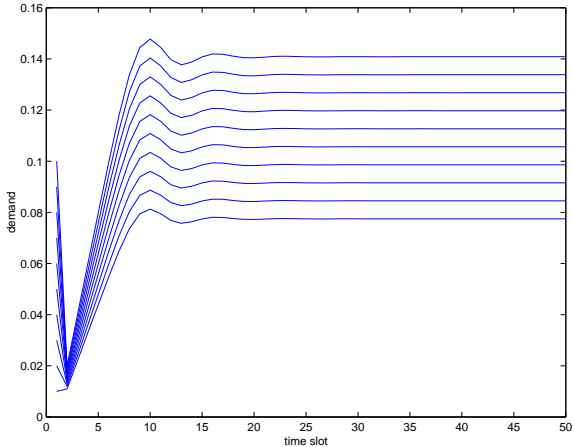}
 \caption{Simulation \ref{s3}: demand adaptation of 10 users}

   \label{demand3}
\end{figure}

\begin{figure}
\centering
\includegraphics[width=3.0in]{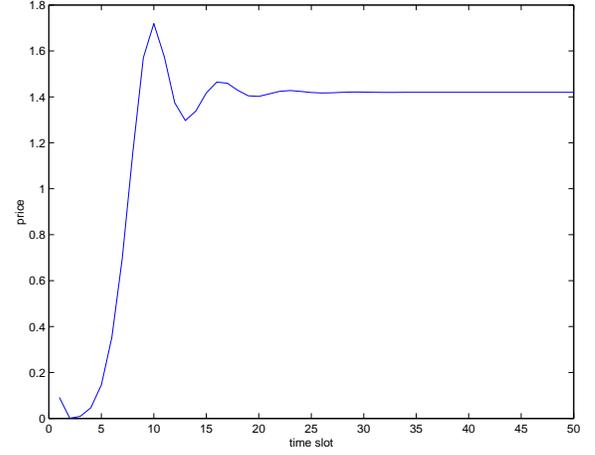}
 \caption{Simulation \ref{s3}: price evolution}
   \label{price3}
\end{figure}

In this simulation experiment we study the effect of heterogeneity
of initial demands, i.e., ten users start with demands ranging from
0.01 to 0.10 respectively. The results are shown in Figure
\ref{demand3} and Figure \ref{price3}. We observe that different
initial conditions do not affect the system stability and
convergence to equilibrium.

\subsection{Heterogeneous initial demands and adaptation rates}
\label{s4}

\begin{figure}
\centering
\includegraphics[width=3.0in]{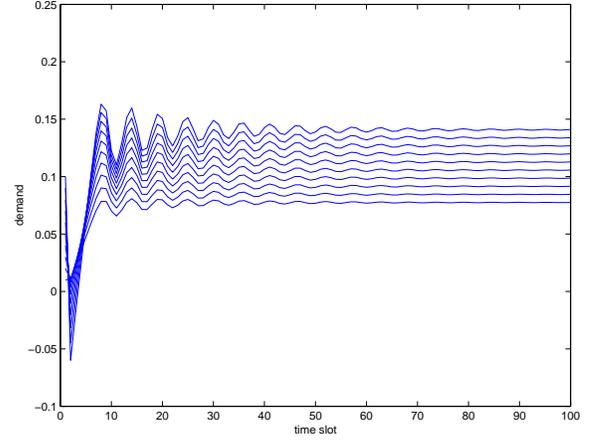}
 \caption{Simulation \ref{s4}: demand adaptation of 10 users}
  \label{demand4}
\end{figure}

\begin{figure}
\centering
\includegraphics[width=3.0in]{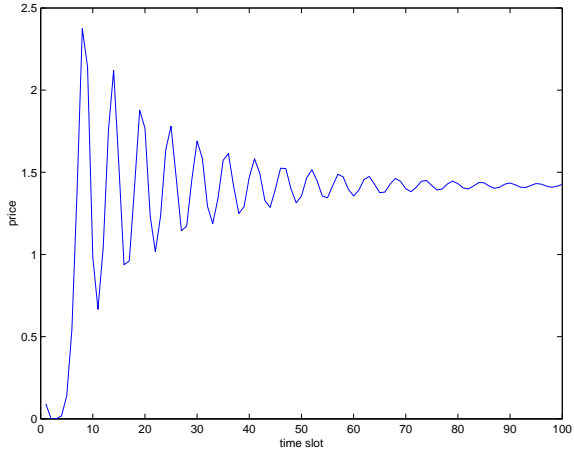}
 \caption{Simulation \ref{s4}: price evolution}
   \label{price4}
\end{figure}

In addition to heterogeneous initial demands as in last simulation,
here users also have different adaptation rates $\alpha_i$: ranging
from 0.11 to 0.20. The results are shown in Figure \ref{demand4} and
Figure \ref{price4}, where we can see that the system still
converges to the equilibrium.

\subsection{Time-varying $w$}
\label{s5}

\begin{figure}
\centering
\includegraphics[width=3.0in]{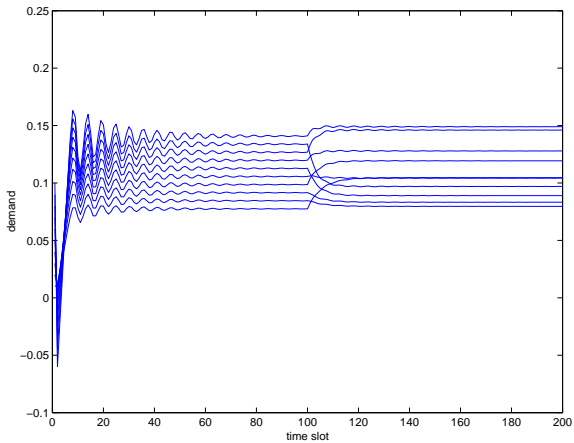}
 \caption{Simulation \ref{s5}: demand adaptation of 10 users}
  \label{demand5}
\end{figure}

\begin{figure}
\centering
\includegraphics[width=3.0in]{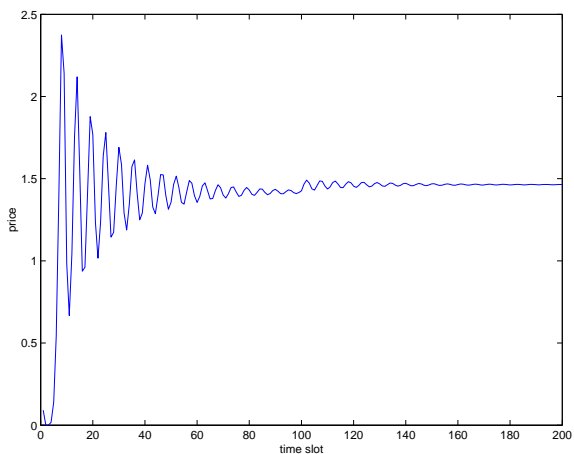}
 \caption{Simulation \ref{s5}: price evolution}
   \label{price5}
\end{figure}

To model the situation where users change their WTP $w$ on-the-fly
to accommodate their energy needs, we change $w_i$'s at time slot
100 by adding a random number within the region of $(-0.05, 0.05)$.
Figure \ref{demand5} and Figure \ref{price5} clearly show that after
time 100, the system tracks the change nicely to a new equilibrium.

\subsection{The effect of $C$} \label{s6}

\begin{figure}
\centering
\includegraphics[width=3.0in]{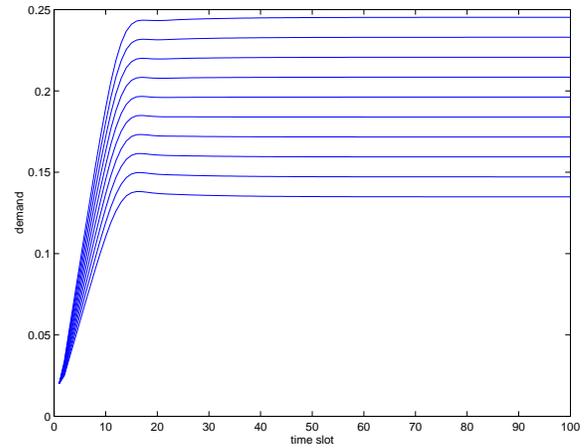}
 \caption{Simulation \ref{s6}: demand adaptation of 10 users}
  \label{demand6}
\end{figure}

\begin{figure}
\centering
\includegraphics[width=3.0in]{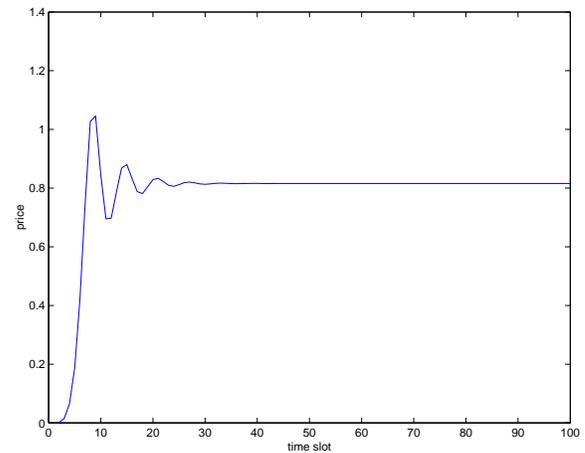}
 \caption{Simulation \ref{s6}: price evolution}
   \label{price6}
\end{figure}

The energy provider can influence the price by adjusting the
capacity $C$. As shown in Figure \ref{demand6} and Figure
\ref{price6}, when $C$ is doubled, the price will drop by $1 - C^{k
\over {k+1}}$ which is $43\%$, and each user's demand will increase
accordingly.

\subsection{Inaccurate price signals} \label{s7}

\begin{figure}
\centering
\includegraphics[width=3.0in]{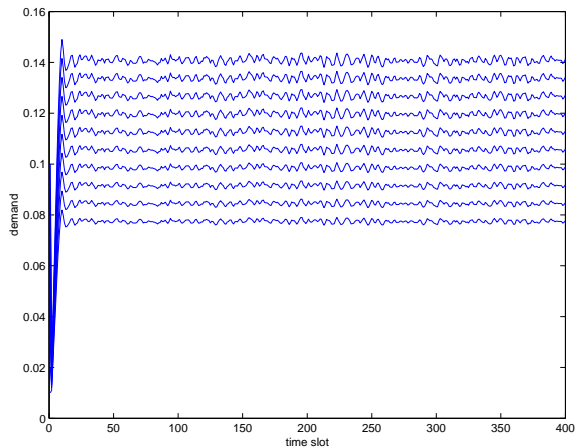}
 \caption{Simulation \ref{s7}: demand adaptation of 10 users}
  \label{demand7}
\end{figure}

\begin{figure}
\centering
\includegraphics[width=3.0in]{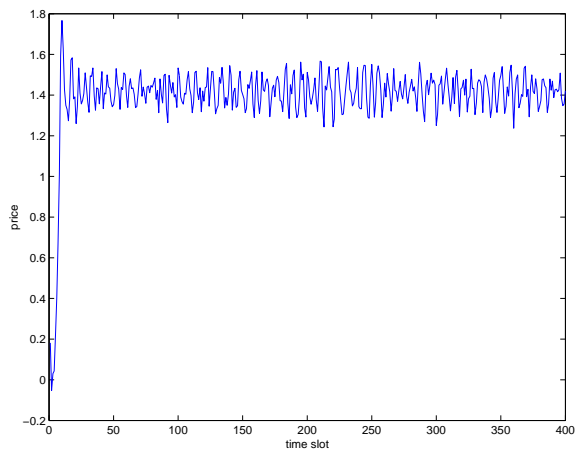}
 \caption{Simulation \ref{s7}: price evolution}
   \label{price7}
\end{figure}

The feedback price signals are transmitted via a communication
network (e.g. GPRS) to the HEM, during which packet loss and delay
could occur. In this case users may have to adapt their demands
based on outdated or inaccurate price information. We model this
situation as a small perturbation to the price signal and study its
effect on the system behavior. As shown in Figure \ref{demand7} and
Figure \ref{price7}, the price and demands still converge to the
means of the equilibrium values, but with small fluctuations. A more
detailed perturbation analysis is part of our future work.

\section{Conclusion and future work}
This paper proposes a distributed framework for demand response and
user adaptation in smart grid networks. More specifically, we have
applied the concept of congestion pricing in Internet traffic
control to the DR problem and shown that it is possible that the
burden of load leveling can be shifted from the grid (or supplier)
to end users via pricing. Individual users adapt to the price
signals to maximize their own benefits. User preference is modeled
as a willingness to pay parameter which can be seen as an indicator
of differential quality of service. The convergence of the algorithm
has been demonstrated by both analysis and simulation results.

This paper is just a first step towards our vision of fully
distributed demand response. There are a number of directions for
future research. Firstly, the proposed model fits nicely into the
game theory framework. In fact, there is already a rich literature
in the networking community on game theoretical analysis of
congestion pricing, e.g. \cite{pbkey}. In Kelly's framework, the
user and social optima coincide if the prices are right, and the
social optimum is a Nash bargaining solution with the logarithmic
utility function. More recently, researchers have applied
intelligent agents and game theory to micro-storage management for
the smart grid \cite{soton}, in which Nash equilibrium is reached
when the agents are able to optimize the energy usage and storage
profile of the dwelling and learn the best storage profile given
market prices at any particular time. Similarly, based on our model,
it would be interesting to study the system dynamics and user
interaction in a large scale energy demand game context. There are
two types of user strategies: price taking users and price
anticipating users \cite{gibbens}. A price taking user assumes that
he has no effect on the price of the energy, whereas a price
anticipating user realizes that his own choice of $w_i$ affects the
price. An interesting observation is that price anticipating users
tend to pay less.

One important element of intelligence in the smart grid is the
learning capability of various components. In demand response, if
users can learn from past observations (e.g. prices and load
profiles), then they can predict the future load and price and
adjust their strategies accordingly (e.g. adjusting $\alpha_i$ and
$w_i$). In this context, Bayesian networks and reinforcement
learning are some of the powerful tools we can leverage to enable
learning in this highly dynamic environment.

As mentioned earlier, demand scheduling can be formulated as a
typical resource allocation problem, for which a wide range of
techniques (many of them have been applied in the networking field)
are available. We are currently investigating the feasibility of
applying some convex optimization techniques such as water-filling
to demand side management, where a certain cost function is to be
minimized subject to a number of system constraints.


%

\section*{Acknowledgment}

The author would like to thank his colleagues at Toshiba Research
Europe for helpful discussions and its Directors for their support
of this work.

\ifCLASSOPTIONcaptionsoff
  \newpage
\fi



\bibliographystyle{IEEEtran}
%

\bibliography{ref2}

\begin{thebibliography}{10}
\providecommand{\url}[1]{#1}
\csname url@samestyle\endcsname
\providecommand{\newblock}{\relax}
\providecommand{\bibinfo}[2]{#2}
\providecommand{\BIBentrySTDinterwordspacing}{\spaceskip=0pt\relax}
\providecommand{\BIBentryALTinterwordstretchfactor}{4}
\providecommand{\BIBentryALTinterwordspacing}{\spaceskip=\fontdimen2\font plus
\BIBentryALTinterwordstretchfactor\fontdimen3\font minus
  \fontdimen4\font\relax}
\providecommand{\BIBforeignlanguage}[2]{{%
\expandafter\ifx\csname l@#1\endcsname\relax
\typeout{** WARNING: IEEEtran.bst: No hyphenation pattern has been}%
\typeout{** loaded for the language `#1'. Using the pattern for}%
\typeout{** the default language instead.}%
\else
\language=\csname l@#1\endcsname
\fi
#2}}
\providecommand{\BIBdecl}{\relax}
\BIBdecl

\bibitem{fan10}
Z.~Fan, G.~Kalogridis, C.~Efthymiou, M.~Sooriyabandara, M.~Serizawa, and
  J.~McGeehan, ``The new frontier of communications research: Smart grid and
  smart metering,'' in \emph{ACM e-Energy}, 2010.

\bibitem{rahimi}
F.~Rahimi and A.~Ipakchi, ``Demand response as a market resource under the
  smart grid paradigm,'' \emph{IEEE Trans. Smart Grid}, Jun. 2010.

\bibitem{wangchen}
C.~Wang and M.~de~Groot, ``Managing end-user preferences in the smart grid,''
  in \emph{ACM e-Energy}, 2010.

\bibitem{wong}
A.~Mohsenian-Rad, V.~Wong, J.~Jatskevich, and R.~Schober, ``Optimal and
  autonomous incentive-based energy consumption scheduling algorithm for smart
  grid,'' in \emph{IEEE PES Conference on Innovative Smart Grid Technologies},
  2010.

\bibitem{kelly}
F.~Kelly, A.~Maulloo, and D.~Tan, ``Rate control for communication networks:
  shadow prices, proportional fairness and stability,'' \emph{Journal of the
  Operational Research Society}, 1998.

\bibitem{glen}
G.~Vinnicombe, ``On the stability of networks operating {TCP}-like congestion
  control,'' in \emph{IFAC}, 2002.

\bibitem{ecn}
S.~Floyd, ``{TCP} and explicit congestion notification,'' \emph{ACM CCR}, Oct.
  1994.

\bibitem{ganesh}
A.~Ganesh, K.~Laevens, and R.~Steinberg, ``Congestion pricing and user
  adaptation,'' in \emph{IEEE Infocom}, 2001.

\bibitem{lemmon}
H.~Bessembinder and M.~Lemmon, ``Equilibrium pricing and optimal hedging in
  electricity forward markets,'' \emph{Journal of Finance}, Jun. 2002.

\bibitem{pbkey}
P.~Key and D.~McAuley, ``Differential {Q}o{S} and pricing in networks: where
  flow control meets game theory,'' \emph{IEE Proceedings Software}, Feb. 1999.

\bibitem{soton}
P.~Vytelingum, T.~Voice, S.~Ramchurn, A.~Rogers, and N.~Jennings, ``Agent-based
  micro-storage management for the smart grid,'' in \emph{AAMAS}, 2010.

\bibitem{gibbens}
R.~Gibbens and F.~Kelly, ``Resource pricing and the evolution of congestion
  control,'' \emph{Automatica}, 1999.

\end{thebibliography}

%

\begin{IEEEbiographynophoto}{Zhong Fan}
is a Research Fellow with Toshiba Research Europe in Bristol, UK. He
received his BSc and MSc degrees in Electronic Engineering from
Tsinghua University, China and his PhD degree in Telecommunication
Networks from Durham University, UK. His research interests are
protocol design and performance analysis of wireless networks, IP
networks, and smart grid communications.
\end{IEEEbiographynophoto}



\vfill


\end{document}